\documentclass[aps,preprint,superscriptaddress,showpacs]{revtex4}%
\usepackage{amsfonts}
\usepackage{amsmath}
\usepackage{amssymb}
\usepackage{graphicx}
\usepackage{graphicx}%
\begin{document}
\title{A single photon emitted by a single particle in free space vacuum modes and
its resonant interaction with two- and three-level absorbers}
\author{R. N. Shakhmuratov}
\affiliation{Instituut voor Kern- en Stralingsfysica, Katholieke
Universiteit Leuven, Celestijnenlaan 200 D, B-3001 Leuven,
Belgium}\affiliation{Optique Nonlin\'{e}aire Th\'{e}orique,
Universit\'{e} Libre de Bruxelles, Campus Plaine CP 231, B-1050
Bruxelles, Belgium}\affiliation{Kazan Physical-Technical
Institute, Russian Academy of Sciences, 10/7 Sibirsky Trakt, Kazan
420029 Russia}
\author{J. Odeurs}
\affiliation{Instituut voor Kern- en Stralingsfysica, Katholieke Universiteit Leuven,
Celestijnenlaan 200 D, B-3001 Leuven, Belgium}
\author{Paul Mandel}
\affiliation{Optique Nonlin\'{e}aire Th\'{e}orique, Universit\'{e} Libre de Bruxelles,
Campus Plaine CP 231, B-1050 Bruxelles, Belgium}
\date{{ \today}}

\begin{abstract}
We consider the time-delayed coincidence counting of two photons emitted in a
cascade by a single particle (atom, molecule, nucleus, etc). The
time-dependence of the probability amplitude of the second photon in the
cascade has a sharply rising leading edge due to the detection of the first
photon, as results from causality. If a macroscopic ensemble of resonant
two-level absorbers is placed in the path of the second photon between the
radiation source and the detector, the photon absorption does not follow
Beer's law due to the time-asymmetric shape of the photon. For very short
delay times almost no absorption takes place, even in an optically dense
medium. We analyze the propagation of such a second photon in a thick resonant
three-level absorber if a narrow electromagnetically induced transparency
(EIT) window is present at the center of the absorption line. It is shown that
the EIT medium can change the asymmetric time dependence of the photon
probability amplitude to a bell shape (EIT filtering). This bell-shaped photon
interacts much more efficiently with an other ensemble of two-level absorbers
chosen, for example, to store this photon and the information it carries.

\end{abstract}
\maketitle

\section{Introduction}

In the last decade, there has been a considerable growth of interest in single
photon experiments in the visible or near infrared domain, mostly related to
information storage, quantum computing and quantum cryptography. An essential
element in these experiments is a single photon source named "photon gun". It
emits one and only one photon when the experimentalist "pulls the trigger".
Faint laser pulses with extremely low mean photon numbers are only an
approximation of single photon pulses. Their state is close to a Glauber state
also containing two and more photons, and in the semiclassical description the
laser pulse is modelled by a Gaussian pulse. Recently, several types of true
single-photon sources were reported, which are based on the laser excitation
of a single trapped ion or atom \cite{1a}, a single organic dye molecule in a
solvent \cite{Moerner},\cite{1m} or a single nitrogen-vacancy color center in
a diamond nanocrystal \cite{1v}. A single-photon turnstile device that uses a
single quantum dot or single quantum well was also proposed \cite{1qd}. Such a
photon is generated spontaneously from a single quantum object (atom, molecule
or quantum dot) that is placed at time $t_{0}=0$ in an excited state by a
short laser pulse or electronically by injecting a single electron and a
single hole to annihilate in a light emitting domain (a central quantum well
in a p-n junction).

Single photon sources based on a single emitting particle can be divided in
two kinds. A source of the first kind radiates in free space vacuum modes. For
example, sources \cite{Moerner},\cite{1m} and \cite{1v} belong to the first
kind. A source of the second kind radiates to a cavity mode. If a high-finesse
cavity is used in a regime of strong coupling in cavity quantum
electrodynamics, almost all radiation is collected in the active cavity mode
and transferred through the cavity mirror (loss channel) in a well defined
direction (see proposals \cite{EberlyLaw},\cite{Kimble} based on the
Jaynes-Commings model \cite{Jaynes}). This process can be made deterministic
and it allows the generation of a single photon with a controlled waveform
\cite{Kuhn},\cite{McKeever},\cite{Walther}. In a quantum network consisting of
spatially separated nodes connected by quantum communication channels it is
preferable to operate with photons having a shape symmetric in time
\cite{Cirac}. In this paper we show that there is an additional physical
argument demanding a time-symmetric photon wave packet if it is supposed to
use such a photon in a atom-field interaction protocol for storage and
retrieval of the radiation state.

We consider a single photon source of the first kind, which intrinsically
produces a photon with asymmetric temporal envelope because of causality. The
resonant interaction of such a photon with an ensemble of two-level absorbers
(atoms, molecules, nuclei, etc.), referred to as a macroscopic absorber, is
analyzed. We show that multiple scattering of the photon in the forward
direction by the resonant particles displays unusual properties. They
originate from the knowledge of the instant of time when the source particle
was prepared in the excited state and then a photon was emitted, i.e., when
the "trigger" of the photon gun was pulled. In the semiclassical language one
can say that the photon wave packet has a particular time envelope having
properties that are very unusual and different from those known for
bell-shaped pulses. We compare the atom-field interaction of such a photon
with that for a photon wave packet with a Gaussian time-envelope. Then, for a
photon source of the first kind we analyze the possibility to reshape the
photon field envelope with the help of electromagnetically induced
transparency (EIT). The reshaped photon time envelope is close to the one with
a Gaussian envelope and its interaction with another absorber (used, for
example, for information storage or quantum computing) becomes the usual one
typical for bell-shaped pulses.

The paper is organized as follows. In Sec. II we discuss the influence of
causality on the spectrum of the single photon emitted by a source of the
first kind and show that it introduces a broad part. In Sec. III we consider
the propagation of such a photon in a thick resonant absorber. In Sec. IV we
study the photon filtering through a narrow EIT window. We show that the
narrow spectrum part of the photon and the broad part are separated in time.
Application of single photon filtering for quantum storage is discussed in
Sec. V. The conditions for level mixing induced transparency for gamma
radiation are considered in Sec. VI.

\section{The influence of causality on the spectrum of a single photon source
of the first kind}

Assume that at time $t_{0}=0$ a single particle (generally a quantum object
such as an atom, a molecule, a nucleus, etc.) is placed in an excited state
$e$. If the radiation, emitted by this particle, has only natural lifetime
broadening due to the coupling with the vacuum modes of free space, the
radiation state is described by the usual expression \cite{2}
\begin{equation}
\left\vert b\right\rangle =\sum_{\mathbf{k}}g_{\mathbf{k}}\frac{\exp
(-i\mathbf{k\cdot r}_{0})}{\nu_{k}+i\Delta_{\text{ph}}}\left\vert
1_{\mathbf{k}}\right\rangle , \label{Eq1a}%
\end{equation}
where $\left\vert 1_{\mathbf{k}}\right\rangle $ is a single-photon Fock state
of the mode $\omega_{k}$ with wave vector $\mathbf{k}$, $\mathbf{r}_{0}$ is
the location of the particle, $\nu_{k}=\omega_{k}-\omega_{0}$ is the frequency
difference of the $k$-mode and the resonant transition from the excited state
$e$ to the ground state $g$, and $2\Delta_{ph}$ is the decay rate of the
excited state $e$. The coupling parameter of the radiation with the source is
$g_{\mathbf{k}}$. This state is actually a wave packet consisting of many Fock
states and normalized such that in total it contains a single photon. Let%
\begin{equation}
E^{(+)}(\mathbf{r},t)=%
{\displaystyle\sum\limits_{\mathbf{k}}}
\widehat{\epsilon}_{\mathbf{k}}\mathcal{E}_{\mathbf{k}}a_{\mathbf{k}}%
e^{-i\nu_{k}t+i\mathbf{k\cdot r}}, \label{Eq1b}%
\end{equation}
be the electric field operator containing only the annihilation operators
$a_{\mathbf{k}}$, $\widehat{\epsilon}_{\mathbf{k}}$ is the unit polarization
vector, $\mathcal{E}_{k}$ is a normalized amplitude of the mode $\mathbf{k}$.
Performing the sum over the wave vector $\mathbf{k}$ in the expression for the
single photon field $b(t)=\left\langle 0\right\vert E^{(+)}(\mathbf{r}%
,t)\left\vert b\right\rangle $, one obtains \cite{2}%
\begin{equation}
b(t)=\frac{\mathcal{E}_{0}}{d}\Theta\left(  t-d/c\right)  e^{-(i\omega
_{0}+\Delta_{\text{ph}})\left(  t-d/c\right)  }, \label{Eq1c}%
\end{equation}
where $d=\left\vert \mathbf{r}-\mathbf{r}_{0}\right\vert $ is the distance
from the source, $\mathcal{E}_{0}$ is a normalized amplitude and $\Theta(t)$
is the Heaviside step function. This wave packet has a sharply rising leading
edge at $t=d/c$ and an exponentially decaying tail. The former is defined by
the time $t_{0}=0$ at which the source is placed in the excited state and the
latter specifies the coherence time or the mean correlation time of the photon
$\tau_{\text{ph}}=1/\Delta_{\text{ph}}$ \cite{2cl}. Such a time dependence of
the single-photon field was detected from radiation of a single organic dye
molecule in solvents \cite{1m}, in time delayed coincidence measurements
(TDCM) of photons emitted in an atomic cascade (optical domain) \cite{2cl} and
in a nuclear cascade (gamma domain) \cite{3n,Hayashi,Balabansky,Haas,3H}. A
typical energy diagram and a detection scheme of TDCM are shown in Fig.
1(a,b). The delayed coincidence counting technique leads to the detailed
observation of the time correlation between successively emitted photons
$a(t)$ and $b(t)$ in the $h\rightarrow e\rightarrow g$ cascade. The photons of
interest, $a(t)$ and $b(t)$, can be selected by interference filters (optical
domain) \cite{2cl} or these photons can be of very different energies such
that different kind of detectors are used to detect them separately (gamma
domain) \cite{3n,Hayashi,Balabansky,Haas,3H}. In the TDCM detection scheme the
energy of these photons is converted by high-speed, high-gain photomultipliers
to electric pulses, which are then amplified and sent to fast discriminators.
A time-to-amplitude convertor is turned on by a pulse due to a $a(t)$ photon
and shut off by one due to a $b(t)$ photon.

Actually it is the probability $P(t)=\left\vert b(t)\right\vert ^{2}$ and not
the complex probability amplitude $b(t)$ which is measured. When a photon is
detected at time $t=t_{n}$, we have a collapse of the photon wave function at
the detector place, which shows as a single count of the detector. After many
events (detector counts), one can reconstruct the actual time distribution of
the probability $P(t)$. One can also extract information about the absolute
value of the probability amplitude since $\left\vert b(t)\right\vert
=\sqrt{P(t)}$. If we know that at time $t_{0}=0$ the source particle is placed
in the excited state $e$ and there is no absorber between the source and the
detector, this probability is zero for $t<0$, it has a maximum, 1, at $t=+0$
and from then on it decreases exponentially. The time resolution of this
measurement is defined by the response time of the detector and electronic
circuit. Therefore, the stepwise rise of the probability $P(t)$ at $t=0$ is
usually slightly smoothened.

For many applications it is important to know how a single photon field
interacts with either a single absorber or an ensemble of quantum absorbers.
In the time domain the propagation of such an asymmetric field through a thick
absorptive medium was considered quantum mechanically \cite{Harris61} and
semiclassically \cite{3n} for gamma photons and for small amplitude pulses in
quantum optics \cite{3o}. It was shown that if the source and the absorber
have the same resonant frequency and the same linewidth, the transmitted
probability amplitude is $b(T,\tau)=\exp(-i\omega_{0}\tau)b_{0}(T,\tau)$,
where%
\begin{equation}
b_{0}(T,\tau)=e^{-\Delta_{\text{ph}}\tau}J_{0}\left(  2\sqrt{T\Delta
_{\text{ph}}\tau}\right)  \Theta(\tau), \label{Eq1d}%
\end{equation}
is the time envelope normalized to 1 for $T=0$ and $\tau=0$, $\tau=t-l/c$ is
the local time, $l$ is the physical length of the macroscopic absorber,
$T=\alpha_{0}l/\gamma$ is its effective thickness, $J_{0}(x)$ is the
zero-order Bessel function, and $\Theta(\tau)$ is the Heaviside step function.
In the definition of $T$ the parameter $\alpha_{0}$ is the resonant absorption
coefficient defined such that $2\alpha_{0}/\gamma$ is Beer's constant, where
$\gamma$ is the halfwidth of the absorption line for the absorber. For this
particular case we take $\gamma=\Delta_{\text{ph}}$. The incident probability
amplitude is assumed to be unity at the front face of the absorber ($l=0$) for
$t=0$. According to Eq. (\ref{Eq1d}) the probability amplitude of the photon
leaving a very thick sample ($T\gg1$) in the forward direction is also close
to unity for the leading pulse edge $\tau\approx0$, the tail of the field
being strongly absorbed and displaying an oscillatory behavior. This was
experimentally detected for gamma photons in
\cite{3n,Hayashi,Balabansky,Haas,3H}. The process was called speed-up of the
initial decay and dynamical beat for longer times (see, for example, the
review \cite{4b}). Thus, the amplitude damping of the output radiation with
increase of $l$ (or $T$) does not follow Beer's law and the time integrated
intensity of this field decreases as $(2\pi T)^{-1/2}$ \cite{3o} instead of
exponential decrease $e^{-2T}$. This is consistent with absorption in
frequency domain of gamma quanta in a thick resonant absorber, which is
calculated for a radiation source with a Lorentzian power spectrum
\cite{MoesBook}.

We argue below that the lack of absorption at $\tau\approx0$ results from the
stepwise form of the initial part of the probability amplitude, i.e., it
results from causality. In a cascade emission of two photons, the time at
which the intermediate excited state $e$ of the source particle is populated
is known by detecting the first photon. From then on, and not earlier, the
second photon can be emitted. Implicit in this scheme is the assumption that
the emission of the first photon projects the atom into the intermediate level
with absolute certainty. In quantum mechanical language, we project the
two-photon state (produced in an atomic or nuclear cascade) onto a single
photon state by detecting the first photon. This detection (the measurement)
introduces a broad part to the otherwise narrow spectrum of the second photon
characterized by the lifetime of the intermediate level $\tau_{\text{ph}}/2$.
A similar process in the photon gun of the first kind, i.e., the pumping of a
single particle by a short laser pulse to the upper state followed by fast
nonradiative decay to a fluorescent state, introduces a broad part to the
spectrum of a single photon emitted in free space vacuum modes. This is the
price paid for pulling the trigger at a desired time, i.e., for the knowledge
of the time at which the excited state is populated. The time distribution of
the detection probability of the photon emitted by the photon gun of the first
kind looks similar to that observed by TDCM of photons emitted in an atomic or
nuclear cascade (compare, for example, the plots for the photon probability
presented in \cite{1m} and \cite{2cl,3n}).

\section{Spectrally narrow and broad parts of a single photon emitted by a
single particle in free space vacuum modes}

In this section we give a brief outline of the methods used to treat the
propagation of a single photon in a resonant absorber and introduce a
decomposition of the photon spectrum (produced by a single particle in free
space) in two components, one being a narrow and the other being broad.

For simplicity we consider the probability amplitude of the emitted photon
$b(t)=b_{0}(t)\exp(-i\omega_{0}t)$ normalized as $b_{0}(t)=\exp(-\Delta
_{\text{ph}}t)\Theta(t)$. The propagation of a single photon in a thick
absorber consisting of particles with the same resonant frequency $\omega_{0}$
and the same lifetime of the excited state as in the emitter is described by
\cite{3n,Harris61,3o}%
\begin{equation}
b_{0}(l,t)=\frac{1}{2\pi}\int_{-\infty}^{+\infty}b_{0}(0,\nu)e^{-i\nu\left(
t-l/c\right)  -A_{eg}(\nu)l}d\nu, \label{Eq21}%
\end{equation}
where
\begin{equation}
b_{0}(0,\nu)=\frac{1}{\Delta_{ph}-i\nu} \label{Eq22}%
\end{equation}
is the Fourier transform of the incident photon $b_{0}(t)$ at the input $l=0$
and%
\begin{equation}
A_{eg}(\nu)=\frac{\alpha_{0}}{\gamma-i\nu} \label{Eq23}%
\end{equation}
is the complex spectral function of the absorber. Here $\gamma$ is a half
width of the absorption line of the individual particle and according to the
imposed condition we have $\gamma=\Delta_{\text{ph}}$. The parameter
$\alpha_{0}=2\pi N_{c}\mu^{2}\omega_{0}/nc$, defined in the previous section,
contains the concentration $N_{c}$ of the resonant particles in the absorber,
the refractive index $n$ of the host in which these particles are
incorporated, and the matrix element $\mu$ of the radiative transition $e-g$.
The expression (\ref{Eq21}) was derived quantum mechanically by Harris
\cite{Harris61}, who considered three-dimensional resonant multiple scattering
of a gamma photon in the forward direction by nuclei residing in a solid
without regular structure of their positions. The same expression, Eq.
(\ref{Eq21}), also follows from the semiclassical approach considering the
frequency dependence of the complex dielectric constant of the absorber
\cite{3n}. Solving the Maxwell-Bloch equations for a small amplitude pulse
with envelope $b_{0}(t)=\exp(-\Delta_{\text{ph}}t)\Theta(t)$ also gives the
same result \cite{3o}. It was shown in Refs. \cite{3n,3o,Harris61} that
performing the integral in Eq. (\ref{Eq21}) leads to Eq. (\ref{Eq1d}).

To clarify the meaning of Eq. (\ref{Eq1d}), we formally represent the Fourier
transform of the single photon field, Eq. (\ref{Eq22}), as $b_{0}(0,\nu
)=b_{s}(0,\nu)+b_{a}(0,\nu)$, where%
\begin{equation}
b_{s}(0,\nu)=\frac{\Delta_{\text{ph}}}{\Delta_{\text{ph}}^{2}+\nu^{2}},
\label{Eq22x}%
\end{equation}%
\begin{equation}
b_{a}(0,\nu)=\frac{i\nu}{\Delta_{\text{ph}}^{2}+\nu^{2}}, \label{Eq23x}%
\end{equation}
are symmetric and antisymmetric parts, respectively. The former has a
Lorentzian shape and the latter resembles the dispersion part of the atomic
response function. Their time domain counterparts are%
\begin{equation}
b_{s}(0,t)=\frac{1}{2}\exp(-\Delta_{\text{ph}}\left\vert t\right\vert ),
\label{Eq24}%
\end{equation}%
\begin{equation}
b_{a}(0,t)=\left\{
\begin{array}
[c]{ccl}%
\frac{1}{2}\exp(-\Delta_{\text{ph}}t) & \quad\text{if\quad} & t>0,\\
0 & \quad\text{if}\quad & t=0,\\
-\frac{1}{2}\exp(\Delta_{\text{ph}}t) & \quad\text{if\quad} & t<0.
\end{array}
\right.  \label{Eq25}%
\end{equation}
The time dependence of $b_{s}(0,t)$ and $b_{a}(0,t)$ is shown in Fig. 2 by the
dashed lines. We calculated the transmission of these components through a
thick resonant absorber at the same conditions as for $b_{0}(T,t)$ with the
help of Eq. (\ref{Eq21}), where the absorber length $l$ is substituted by its
effective thickness $T=\alpha_{0}l/\gamma$ and $b_{0}(0,\nu)$ is substituted
by $b_{s}(0,\nu)$ or $b_{a}(0,\nu)$. The result is%
\begin{equation}
b_{s}(T,\tau)=\left\{
\begin{array}
[c]{ccl}%
\frac{1}{2}e^{-\Delta_{\text{ph}}\tau}F_{-}(T,\tau) & \quad\text{if}\quad &
\tau\geq0,\\
\frac{1}{2}e^{\Delta_{\text{ph}}\tau-T/2} & \quad\text{if}\quad & \tau<0,
\end{array}
\right.  \label{Eq26}%
\end{equation}%
\begin{equation}
b_{a}(T,\tau)=\left\{
\begin{array}
[c]{ccl}%
\frac{1}{2}e^{-\Delta_{\text{ph}}\tau}F_{+}(T,\tau) & \quad\text{if}\quad &
\tau>0,\\
\frac{1}{2}\left(  1-e^{-T/2}\right)  & \quad\text{if}\quad & \tau=0,\\
-\frac{1}{2}e^{\Delta_{\text{ph}}\tau-T/2} & \quad\text{if}\quad & \tau<0,
\end{array}
\right.  \label{Eq27}%
\end{equation}
where%
\begin{equation}
F_{\pm}(T,\tau)=J_{0}(2\sqrt{T\Delta_{\text{ph}}\tau})\pm\frac{1}{2}%
{\displaystyle\int\nolimits_{0}^{T}}
e^{-(T-x)/2}J_{0}(2\sqrt{x\Delta_{\text{ph}}\tau})dx, \label{Eq28}%
\end{equation}
and $\tau=t-l/c$ is the local time at the output of the absorber of the
physical length $l$. From these expressions one can find that for $\tau=\pm0$
the probability amplitudes take values $b_{s}(T,\pm0)=\exp(-T/2)/2$,
$b_{a}(T,-0)=-\exp(-T/2)/2$, which are much smaller than 1 for large thickness
($T\gg1$), and $b_{a}(T,+0)=1-\exp(-T/2)/2$, which is almost equal to the
probability amplitude of the input radiation. For longer times ($\tau>0$), the
behavior of the probability amplitude of the antisymmetric part $b_{a}%
(T,\tau)$ is close to the total amplitude $b_{0}(T,\tau)$, Eq. (\ref{Eq1d}).
The time evolution of the output probability amplitudes $b_{s}(T,\tau)$ and
$b_{a}(T,\tau)$ for $T=10$ is shown in Fig. 2 by the solid lines, (a) and (b)
respectively. These plots clearly demonstrate that at the output of a thick
sample the probability amplitude of the antisymmetric part becomes very close
to the total probability amplitude of a single photon $b_{0}(T,\tau)$. We may
conclude, that the narrow spectrum part of the photon is strongly absorbed in
a thick sample while the broad component partially passes through and it
defines the probability amplitude of the output radiation. The slight
deviation of $b_{a}(T,\tau)$ from $b_{0}(T,\tau)$ is due to the reduced
absorption of the far wings of the symmetric part of the photon spectrum in
the case of $\Delta_{\text{ph}}=\gamma$. Its time domain counterpart
$b_{s}(T,\tau)$ shows an oscillatory deviation from the exponential behavior
$\exp(-\Delta_{\text{ph}}\left\vert \tau\right\vert -T/2)/2$ for $\tau>0$. To
be convinced that this deviation is due to the reduced absorption of the wings
of the symmetric part of the photon spectrum, we consider the photon
propagation through a medium with particles whose absorption line $\Gamma$ is
much broader than the spectral width of the input photon $\Delta_{\text{ph}}$.
In this case we expect that the contribution of the symmetric part to the
total output amplitude $b_{0}(T,\tau)$ is negligible since it must be strongly absorbed.

In Ref. \cite{3o} it was shown that if, for example, we have inhomogeneous
broadening of the absorption line in a macroscopic absorber, the output
radiation field is described by Eq. (\ref{Eq21}) where the spectral function
$A_{eg}(\nu)$ is substituted by%
\begin{equation}
A_{\Gamma}(\nu)=\frac{\alpha_{0}}{\Gamma-i\nu} \label{Eq29}%
\end{equation}
with a total halfwidth $\Gamma=\gamma+\gamma_{inh}$, where $\gamma_{inh}$ is
inhomogeneous halfwidth. Generally, with the help of the linear response
approximation, one can derive the same expression for a weak radiation field
transmitted through a resonant absorber having an absorption halfwidth
$\Gamma$ (natural, homogeneously or inhomogeneously broadened by the neighbors
of the resonant particles in the absorber), which is different from the
spectral halfwidth of the incoming radiation $\Delta_{\text{ph}}$
($\Delta_{\text{ph}}\ll\Gamma$). We calculated the transmission of the
probability amplitudes of the symmetric and antisymmetric parts of incoming
radiation for such a sample with spectral function $A_{\Gamma}(\nu)$. The
result is%
\begin{equation}
b_{s}(l,\tau)=\left\{
\begin{array}
[c]{rcl}%
\frac{1}{2}e^{-\Delta_{\text{ph}}\tau-T_{-}}+f_{-}(l,\tau) & \quad
\text{if}\quad & \tau\geq0,\\
\frac{1}{2}e^{\Delta_{\text{ph}}\tau-T_{+}} & \quad\text{if}\quad & \tau<0,
\end{array}
\right.  \label{Eq30}%
\end{equation}%
\begin{equation}
b_{a}(l,\tau)=\left\{
\begin{array}
[c]{rcl}%
\frac{1}{2}e^{-\Delta_{\text{ph}}\tau-T_{-}}+f_{+}(l,\tau) & \quad
\text{if}\quad & \tau>0,\\
\frac{1}{2}\left(  1-e^{-T_{+}}\right)  & \quad\text{if}\quad & \tau=0,\\
-\frac{1}{2}e^{\Delta_{\text{ph}}\tau-T_{+}} & \quad\text{if}\quad & \tau<0,
\end{array}
\right.  \label{Eq31}%
\end{equation}
where%
\begin{equation}
f_{\pm}(l,\tau)=\frac{e^{-\Gamma\tau}}{2}\left[  g(T_{-},\tau)\pm g(T_{+}%
,\tau)\right]  , \label{Eq32}%
\end{equation}%
\begin{equation}
g(T_{\pm},\tau)=%
{\displaystyle\int\nolimits_{0}^{T_{\pm}}}
e^{-(T_{\pm}-x)}J_{0}(2\sqrt{x(\Gamma\pm\Delta_{\text{ph}})\tau})dx.
\label{Eq33}%
\end{equation}
Here $T_{\pm}=\alpha_{0}l/(\Gamma\pm\Delta_{\text{ph}})$ is the effective
thickness of the absorber. Formally, with the decrease of $\Delta_{\text{ph}}$
($\Delta_{\text{ph}}\rightarrow0$) the function $f_{-}(l,\tau)$ in the
solution $b_{s}(l,\tau)$ tends to zero since its components $g(T_{+},\tau)$
and $g(T_{-},\tau)$ nearly cancel each other. However, their difference is
larger than the first term $(1/2)\exp(-\Delta_{\text{ph}}\tau-T_{-})$ in
$b_{s}(l,\tau)$ for $\tau>0$, and, hence, we have residual oscillations of the
symmetric part of the photon amplitude for $\tau>0$. In spite of this, for
$\tau>0$, it is the antisymmetric part $b_{a}(l,\tau)$ that gives the dominant
contribution to the total probability amplitude of the output photon. This
amplitude, $b_{0}(l,\tau)$, can be approximated by the two main terms obtained
from the successive integration by parts of the integrals in Eqs.
(\ref{Eq30}),(\ref{Eq31}), which are
\begin{equation}
b_{0}(l,\tau)=e^{-\Gamma\tau}\left[  J_{0}(2\sqrt{\alpha_{0}l\tau}%
)+(\Gamma-\Delta_{\text{ph}})\tau\frac{J_{1}(2\sqrt{\alpha_{0}l\tau})}%
{\sqrt{\alpha_{0}l\tau}}\right]  \Theta(\tau). \label{Eq34}%
\end{equation}
$J_{1}(x)$ is the Bessel function of the first order. It is remarkable that
the first and also main term of this expression does not depend on the
spectral width $\Delta_{\text{ph}}$ of the incoming photon. Thus, for
$\Delta_{\text{ph}}\ll\Gamma$ the amplitude of the output radiation is mainly
defined by the parameters of the sample and not those characterizing the input photon.

The symmetric part of the input photon spectrum has a Lorentzian shape with
halfwidth $\Delta_{\text{ph}}$ that is narrow with respect to the absorption
linewidth, i.e., $\Delta_{\text{ph}}\ll\Gamma$. Therefore, this part is
strongly absorbed in a thick sample. The antisymmetric part partially passes
through a thick sample and what comes out is due to the far wings of
$b_{a}(0,\nu)$, which are proportional to $1/\nu$ and independent on
$\Delta_{\text{ph}}$. This is the reason why the output radiation does not
contain information about $\Delta_{\text{ph}}$. Therefore, formally we refer
to this part of the photon spectrum as broad. Fig. 3(a) shows a comparison of
the time dependence of the antisymmetric part $b_{a}(l,\tau)$ with the
numerically calculated integral (\ref{Eq21}) for the total amplitude
$b_{0}(l,\tau)$, where the spectral function is $A_{\Gamma}(\nu)$. These plots
show that the probability amplitude of the output radiation essentially
coincides with the probability amplitude of the antisymmetric part
$b_{a}(l,\tau)$, confirming the heuristic argument that the symmetric part is
strongly absorbed.

Another argument supporting the concept of the broad spectrum component of the
photon can be given with the help of the time integrated intensity of the
output radiation. For a classical field, the total energy transmitted through
a unit area at distance $l$ is proportional to%
\begin{equation}
U(l)=%
{\displaystyle\int\nolimits_{-\infty}^{+\infty}}
b(l,\tau)b^{\ast}(l,\tau)d\tau. \label{Eq35}%
\end{equation}
For a single photon, this value is actually proportional to the number of
counts of the second detector in a wide time window obtained without TDCM,
i.e., without the first detector in the scheme Fig. 1(b). Eq. (\ref{Eq35}) can
be transformed as%
\begin{equation}
U(l)=\frac{1}{2\pi}%
{\displaystyle\int\nolimits_{-\infty}^{+\infty}}
b(l,\nu)b^{\ast}(l,\nu)d\nu, \label{Eq36}%
\end{equation}
where $b(l,\nu)=b(0,\nu)\exp[-A_{m}(\nu)l]$ and $A_{m}(\nu)$ is $A_{eg}(\nu)$
or $A_{\Gamma}(\nu)$.

In the case of a macroscopic absorber with a narrow spectral function
$A_{eg}(\nu)$ of halfwidth $\Delta_{\text{ph}}$, we have for the symmetric,
$b_{s}(l,\nu)$, and antisymmetric, $b_{a}(l,\nu)$, parts of the output
photon:
\begin{equation}
U_{s}(T)=e^{-T}[I_{0}(T)-I_{1}(T)]U_{0}(0)/2, \label{Eq37}%
\end{equation}%
\begin{equation}
U_{a}(T)=e^{-T}[I_{0}(T)+I_{1}(T)]U_{0}(0)/2, \label{Eq38}%
\end{equation}
where $U_{0}(0)=\tau_{\text{ph}}/2$ is the time integrated intensity of the
input field $b_{0}(0,\tau)$, $T=\alpha_{0}l/\Delta_{\text{ph}}$, $I_{0}(T)$
and $I_{1}(T)$ are the modified Bessel functions of zero and first order,
respectively. For the time integrated intensity of the total field
$b_{0}(l,\nu)=b_{s}(l,\nu)+b_{a}(l,\nu)$, we have the expression
$U_{0}(T)=U_{s}(T)+U_{a}(T)$, which is consistent with $U_{0}(T)=\exp
(-T)I_{0}(T)U_{0}(0)$ found in Ref. \cite{3o}.

For an absorber with a broad spectral function $A_{\Gamma}(\nu)$ with a
halfwidth $\Gamma>\Delta_{\text{ph}}$, we have%
\begin{equation}
U_{s}(T_{b})=U_{+}(T_{b})-\left(  \frac{\Delta_{\text{ph}}}{\Gamma}\right)
^{3}\left[  U_{1}(T_{b})-U_{2}(T_{b})\right]  , \label{Eq39}%
\end{equation}%
\begin{equation}
U_{a}(T_{b})=U_{-}(T_{b})+\frac{\Delta_{\text{ph}}}{\Gamma}U_{1}%
(T_{b})-\left(  \frac{\Delta_{\text{ph}}}{\Gamma}\right)  ^{3}U_{2}(T_{b}),
\label{Eq40}%
\end{equation}
where%
\begin{equation}
U_{\pm}(T_{b})=e^{-2aT_{b}}\left[  1\pm4a^{2}\left(  \frac{\Delta_{\text{ph}}%
}{\Gamma}\right)  ^{2}T_{b}\right]  \frac{U_{0}(0)}{2}, \label{Eq41}%
\end{equation}%
\begin{equation}
U_{1}(T_{b})=2a^{2}U_{0}(0)%
{\displaystyle\int\nolimits_{0}^{T_{b}}}
e^{-2a(T_{b}-x)-x}I_{0}(x)dx, \label{Eq42}%
\end{equation}%
\begin{equation}
U_{2}(T_{b})=4a^{3}U_{0}(0)%
{\displaystyle\int\nolimits_{0}^{T_{b}}}
(T_{b}-x)e^{-2a(T_{b}-x)-x}I_{0}(x)dx, \label{Eq43}%
\end{equation}
$a=\Gamma^{2}/(\Gamma^{2}-\Delta_{\text{ph}}^{2})$ and $T_{b}=\alpha
_{0}l/\Gamma$ is the effective thickness of the absorber at the center of the
broad absorption line. Fig. \ref{fig:3}(b) shows the thickness dependence of
the time integrated intensity of the symmetric, $U_{s}(T_{b})$, (dash-dot
line) and antisymmetric, $U_{a}(T_{b})$, (solid line) parts of a single
photon. The antisymmetric part does not follow Beer's law, while the symmetric
part follows this law up to a certain thickness ($T_{b}\lesssim5$) and its
contribution to $U_{0}(T)$ becomes negligible for $T\gtrsim1.5$. For an
absorber with a narrow absorption line of a halfwidth $\Delta_{\text{ph}}$,
both parts (not shown on the plots) do not follow Beer's law for any distance
and their $T$-dependence resembles that for $U_{a}(T_{b})$ shown by thick
solid line in Fig. 3(b).

If $A_{m}(\nu)=A_{eg}(\nu)$, Eq. (\ref{Eq36}) for $U(l)$ coincides with a
formula for the transmission of gamma-photons through a thick absorber of
length $l$ (exact resonance case if the central frequency of the photon equals
the resonant frequency of an absorber), well known in frequency domain
M\"{o}ssbauer spectroscopy \cite{MoesBook}. This means that the Lorentzian
power spectrum of gamma photons, adopted in the general formula
\cite{MoesBook}, implies causality, since the relevant spectrum of the photon
probability amplitude must satisfy Eq. (\ref{Eq1a}) or Eq. (\ref{Eq22}), both
assuming a time-asymmetric photon envelope. If the photon spectrum would be
defined only by the symmetric part $b_{s}(0,\nu)$, Eq. (\ref{Eq22x}), the
power spectrum of such a photon is Lorentzian squared and all peculiarities
mentioned above (violation of Beer's law etc.) would not be present.

Summarizing we conclude that the spectrum of the photon emitted in free space
consists of two parts: a narrow and a broad part. The narrow part comes from
the symmetric component $b_{s}(0,t)$ of the photon probability amplitude. The
amplitude of this part decays with propagation distance and its absorption
follows Beer's law in a medium with a wide absorption line. In a medium with a
narrow absorption line, comparable with the spectral width of the symmetric
part, this component is also strongly absorbed. But due to the reduced
absorption of its far spectrum wings, the frequency integrated absorption of
the symmetric component does not follow Beer's law. The antisymmetric part of
the photon always violates Beer's law, its amplitude at $\tau=+0$ is nearly 1
for any thick absorber, then this part decays with the rate defined by the
spectral width of the absorber and by its effective thickness.

In the next Section we propose to filter out the broad part of the photon
spectrum with the help of electromagnetically induced transparency. The same
mechanism can also reduce the intrinsic spectrum width of the source photon,
i.e., the width of the narrow part.

\section{Photon filtering through the EIT window}

Electromagnetically induced transparency (EIT) allows transmission of the
radiation field through a thick sample without absorption and with appreciable
delay due to the reduced group velocity Ref. \cite{Harris97}. Recently it was
shown that the EIT window allows also a separation of the narrow and broad
components of the radiation field in time \cite{ShOd05}. Here we apply these
results to separate the narrow and broad components of a single photon emitted
in free space. For simplicity we consider a resonant absorber described by a
three-level scheme with a DC coupling field shown in Fig. 4. This scheme was
experimentally realized for atomic hydrogen whose excited electron states $2p$
and $2s$ (degenerate in energy) are coupled by an electric DC field
\cite{Hakuta}. State $2s$ is a metastable state with a long lifetime.
Therefore, the coupling creates a narrow transparency window at the center of
the absorption line.

For a weak input radiation field being in exact resonance with $e-g$
transition, the output field from the absorber with EIT window is described by
Eq. (\ref{Eq21}) with the spectral function $A_{eg}(\nu)$ replaced by
\cite{ShOd05}%
\begin{equation}
A_{\text{eit}}(\nu)=\frac{\alpha_{0}(\gamma_{m}-i\nu)}{(\Gamma-i\nu
)(\gamma_{m}-i\nu)+\Omega^{2}}, \label{Eq44}%
\end{equation}
where $\Gamma$ is the halfwidth of the unperturbed absorption line for the
transition from the ground state $g$ to the excited state $e$, $\Omega$ is the
coupling parameter defined by the strength of the DC field, which couples the
excited state $e$ with the metastable state $m$ (see Fig. 4), and $\gamma_{m}$
is the half decay rate of the metastable state $m$. The EIT hole is deep if
the coupling parameter satisfies the condition $\Omega^{2}\gg\gamma_{m}\Gamma$
and its halfwidth is approximately $\Delta_{\text{eit}}=\Omega^{2}/\Gamma$.

Following Ref. \cite{ShOd05}, we apply the adiabatic expansion of the spectral
function $A_{\text{eit}}(\nu)$ near the center of the EIT hole, $\nu=0$, for
the approximate calculation of the integral in Eq. (\ref{Eq21}) describing the
output photon probability amplitude. This method is valid if the EIT hole is
present at the line center, i.e., if the condition $\Omega^{2}\gtrsim
\gamma_{m}\Gamma$ is satisfied. The method gives a nice approximation of the
spectrally narrow, adiabatic part of the output radiation filtered through the
EIT window in a thick absorber. The nonadiabatic, spectrally broad part of the
output radiation will be treated separately, as in \cite{ShOd05}. In the
adiabatic expansion it is sufficient to retain the three main terms (see
Section III in Ref. \cite{ShOd05}):
\begin{equation}
A_{\text{eit}}(\nu)l\approx T_{\text{eit}}-i\nu t_{d}+\nu^{2}/\Delta
_{\text{eff}}^{2}, \label{Eq6}%
\end{equation}
containing three important parameters of the EIT medium, i.e.,
\begin{equation}
T_{\text{eit}}=T_{b}\frac{\gamma_{m}\Gamma}{\Omega^{2}+\gamma_{m}\Gamma},
\label{Eq7b}%
\end{equation}%
\begin{equation}
t_{d}=T_{b}\Gamma\frac{\Omega^{2}-\gamma_{m}^{2}}{(\Omega^{2}+\gamma_{m}%
\Gamma)^{2}}, \label{Eq7c}%
\end{equation}%
\begin{equation}
\Delta_{\text{eff}}=\sqrt{\frac{(\Omega^{2}+\gamma_{m}\Gamma)^{3}}{T_{b}%
\Gamma\lbrack\Omega^{2}(\Gamma+2\gamma_{m})-\gamma_{m}^{3}]}}. \label{Eq7d}%
\end{equation}
The first parameter, $T_{\text{eit}}$, is the EIT reduced effective thickness
of the absorber, which is defined by the residual absorption at the bottom of
the EIT hole ($\nu=0$). Here, we define the effective thickness $T_{b}$ with
respect to the absorption without the coupling $\Omega$, i.e., $T_{b}%
=\alpha_{0}l/\Gamma$. We have $T_{\text{eit}}/T_{b}\approx\gamma_{m}%
/\Delta_{eit}\ll1$ if $\Omega^{2}\gg\gamma_{m}\Gamma$. The second parameter,
$t_{d}$, approximated as $t_{d}\approx T_{b}/\Delta_{\text{eit}}$, is the
delay time of the radiation field associated with the slow group velocity
$V_{g}=c/(1+ct_{d}/l)$. It is defined by the slope of the steep dispersion at
the center of the EIT window. The third parameter, $\Delta_{\text{eff}}$, is
the effective width of the EIT window for a thick sample, which narrows with
distance as $\sim1/\sqrt{T_{b}}$. This parameter is approximated as
$\Delta_{\text{eff}}$ $\approx\Delta_{\text{eit}}/\sqrt{T_{b}}$ and it is
responsible for the pulse broadening in time or its spectrum narrowing with
distance. With $A_{\text{eit}}(\nu)$, given by Eq. (\ref{Eq6}), the integral
in Eq.(\ref{Eq21}), which can be considered as the inverse Fourier transform,
is calculated with the help of the convolution theorem. This gives an integral
representation of the adiabatic solution for the probability amplitude of the
output photon
\begin{equation}
b_{0A}(l,\tau)=\frac{\Delta_{\text{eff}}}{2\sqrt{\pi}}\int_{-\infty}^{+\infty
}b_{01}(l,\tau_{1}-t_{d})e^{-\frac{1}{4}\Delta_{\text{eff}}^{2}(\tau-\tau
_{1})^{2}}d\tau_{1}. \label{Eq7}%
\end{equation}
The exponential part of the integrand takes into account the third term of the
adiabatic expansion (Eq. (\ref{Eq6})). The other part of the integrand,
$b_{01}(l,\tau_{1}-t_{d})=b_{0}(0,\tau_{1}-t_{d})\exp(-T_{eit})$, is the
probability amplitude obtained using only the first two terms of the adiabatic
expansion, i.e., $A_{\text{eit}}(\nu)l\approx T_{\text{eit}}-i\nu t_{d}$.
Explicitly this function is%
\begin{equation}
b_{01}(l,\tau-t_{d})=e^{-\Delta_{\text{ph}}(\tau-t_{d})-T_{\text{eit}}}%
\Theta(\tau-t_{d}). \label{Eq7a}%
\end{equation}
The integral in Eq. (\ref{Eq7}) can be calculated and gives%
\begin{equation}
b_{0A}(l,\tau)=\phi_{+}(l,\tau)e^{-T_{\text{eit}}-\Delta_{\text{ph}}%
(\tau-t_{d})}, \label{Eq8}%
\end{equation}
where%
\begin{equation}
\phi_{\pm}(l,\tau)=\frac{1}{2}e^{\Delta_{\text{ph}}^{2}/\Delta_{\text{eff}%
}^{2}}\left[  1\pm\operatorname{erf}\left(  \frac{\Delta_{\text{eff}}}{2}%
(\tau-t_{d})\mp\frac{\Delta_{\text{ph}}}{\Delta_{\text{eff}}}\right)  \right]
, \label{Eq9}%
\end{equation}
$\operatorname{erf}(x)$ is the Error function, which is defined as an odd
function, i.e., $\operatorname{erf}(-x)=-\operatorname{erf}(x)$. The function
$\phi_{-}(l,\tau)$ will be used to describe the probability amplitude of the
symmetric and antisymmetric parts of the photon.

If $\Delta_{\text{ph}}\ll\Delta_{\text{eff}}$, the function $\phi_{+}(l,\tau)$
is almost unity for $\tau\gg t_{d}+2\Delta_{\text{eff}}^{-1}$, and it is
almost zero if $\tau\ll t_{d}-2\Delta_{\text{eff}}^{-1}$. In the transient
domain, $t_{d}-2\Delta_{\text{eff}}^{-1}\lesssim\tau\lesssim t_{d}%
+2\Delta_{\text{eff}}^{-1}$, the function $\phi_{+}(l,\tau)$ gradually
increases from 0 to 1. Fig. 5 shows the time dependence of the function
$\phi_{+}(l,\tau)$ for $\Delta_{\text{ph}}/\Delta_{\text{eff}}=0.1$ (solid
line) and for $\Delta_{\text{ph}}/\Delta_{\text{eff}}=0$ (dotted line). We see
that if the spectral width of the narrow part of the photon, $2\Delta
_{\text{ph}}$, is much smaller than the effective width, $2\Delta_{\text{eff}%
}$, of the EIT window for a thick sample ($T_{b}\gg1$), the exponential tail
of the delayed photon is almost not affected by the EIT hole, while the front
edge is smoothened and spread around $t_{d}$ in the domain $t_{d}%
-2\Delta_{\text{eff}}^{-1}<\tau<t_{d}+2\Delta_{\text{eff}}^{-1}$. Thus, the
narrow part is not affected in this case while the broad part is filtered out.
Actually the broad part is present in the output signal, but it is not
delayed, as will be shown below.

If the spectral width of the narrow part of the photon is much larger than the
transparency window, $\Delta_{\text{ph}}\gg\Delta_{\text{eff}}$, then the
function $b_{01}(l,\tau_{1}-t_{d})$ in Eq. (\ref{Eq7}) can be approximated by
a delta function, $\delta(\tau_{1}-t_{d})$, and its convolution with the
Gaussian function is $b_{0A}(l,\tau)\sim\exp\left[  -\frac{1}{4}%
\Delta_{\text{eff}}^{2}(\tau-t_{d})^{2}\right]  $. Thus, for $\Delta
_{\text{ph}}\gg\Delta_{\text{eff}}$ the photon envelope will spread in time
and narrow in spectrum acquiring a Gaussian shape with a temporal width
$4/\Delta_{\text{eff}}$.

Such time-broadening of the pulse almost conserves its area%
\begin{equation}
\theta_{A}(l)=\int_{-\infty}^{+\infty}2\mu b_{0A}(l,\tau)dt=\theta
_{A}(0)e^{-T_{\text{eit}}}, \label{Eq10}%
\end{equation}
whose value is affected only by the residual absorption defined by
$T_{\text{eit}}$, which can be very small ($T_{\text{eit}}\ll1$). Here
$\theta_{A}(0)=2\mu/\Delta_{\text{ph}}$ is the area of the input pulse
$b_{0}(0,t)$. The pulse area is a quantitative parameter specifying the field
interaction with a two-level atom. For example, the probability amplitudes of
ground and excited states of a two-level atom interacting with a short
resonant pulse (shorter than any relaxation time) are given by $\cos
(\theta/2)$ and $i\sin(\theta/2)$ respectively, where $\theta$ is a pulse area
\cite{Eberly}.

The time integrated value of the pulse intensity $\sim\left\vert b_{0A}%
(l,\tau)\right\vert ^{2}$, which is essentially the average detection
probability of the photon,
\begin{equation}
U_{A}(l)=\int_{-\infty}^{+\infty}\left\vert b_{0A}(l,\tau)\right\vert
^{2}d\tau\label{Eq10a}%
\end{equation}
reduces with distance due to broadening of its envelope as
\begin{equation}
U_{A}(l)=U_{0}(0)e^{-2T_{\text{eit}}+2\Delta_{\text{ph}}^{2}/\Delta
_{\text{eff}}^{2}}\left[  1-\operatorname{erf}\left(  \sqrt{2}\Delta
_{\text{ph}}/\Delta_{\text{eff}}\right)  \right]  . \label{Eq11}%
\end{equation}
If $\Delta_{\text{ph}}\ll\Delta_{\text{eff}}/\sqrt{2}$, the output energy does
not change appreciably due to the pulse broadening in time. If $\Delta
_{\text{ph}}\gg\Delta_{\text{eff}}/\sqrt{2}$, the energy is reduced by a
factor that is the ratio of the EIT hole width and the spectral width of the
photon $\sim\Delta_{\text{eff}}/\Delta_{\text{ph}}$.

So far, we have discussed the propagation of the adiabatic component of the
probability amplitude of the photon whose spectrum is confined in the EIT
window. This is the spectrally narrow part. The spectrally broad part is not
sensitive to the narrow frequency domain of the absorption spectrum around
$\nu=0$, where the EIT hole is located. The time evolution of this spectrally
broad part of the photon in a thick resonant absorber is mostly defined by the
far wings of the absorption line, which decrease as $1/\nu^{2}$, and by the
dispersion component, which drops even more slowly $\sim1/\nu$. Therefore, to
describe the propagation of the spectrally broad part we neglect in the
function $A(\nu)$ the coupling $\Omega$, which produces the EIT hole and does
not affect the wings of the absorption line. Then, the broad, nonadiabatic
component of the photon probability amplitude, $b_{0}(l,\tau)$,\ is governed
by the integral (\ref{Eq21}) with the spectral function $A_{\Gamma}(\nu)$
given by Eq. (\ref{Eq29}).

We consider two cases. If $\Delta_{ph}=\Gamma$, we have the case of equal
spectral widths of the incoming photon and of the absorption line for a single
particle. Then the nonadiabatic broad part is $b_{N}(l,\tau)=b_{0N}%
(l,\tau)\exp(-i\omega_{0}\tau)$, where $b_{0N}(l,\tau)$ is described by Eq.
(\ref{Eq1d}). The second case corresponds to the situation where the photon
spectral width is smaller than the absorption linewidth ($\Delta_{ph}<\Gamma
)$. Then $b_{0N}(l,\tau)$ is given by the sum of $b_{s}(l,\tau)$, Eq.
(\ref{Eq30}),\ and $b_{a}(l,\tau)$, Eq. (\ref{Eq31}), or by the simplified
expression (\ref{Eq34}). In both cases the output probability amplitude of the
photon is the sum of the adiabatic (spectrally narrow) and non-adiabatic
(spectrally broad) amplitudes
\begin{equation}
b_{\text{tot}}(l,\tau)=b_{0A}(l,\tau)+b_{0N}(l,\tau). \label{Eq12}%
\end{equation}

First we consider the case $\Delta_{\text{ph}}=\gamma_{m}$. Then the two
terms, $\Delta_{\text{ph}}t_{d}$ and $T_{\text{eit}}$, are almost equal and
they nearly cancel each other in Eq. (\ref{Eq8}), reducing the adiabatic part
of the photon amplitude $b_{0A}(l,\tau)$ to%
\begin{equation}
b_{0A}(l,\tau)\approx\phi_{+}(l,\tau)e^{-\Delta_{\text{ph}}\tau}. \label{Eq13}%
\end{equation}
Figure 6(a) shows the time dependence of the photon probability amplitude
$b_{\text{tot}}(l,\tau)$ (bold line). This is our analytical solution, Eq.
(\ref{Eq12}). It coincides almost perfectly with the numerical evaluation of
Eq. (\ref{Eq21}) with $A_{\text{eit}}(\nu)$, which is an exact expression. The
effective thickness is taken as $T=30$, the other parameters are
$\Gamma=10\gamma_{m}$ and $\Omega=2\Gamma$. The probability amplitude of the
photon at the same distance with no resonant absorber between the source and
the detector, $b_{0}(0,\tau)$, is shown for comparison (thin line). The dashed
line shows the probability amplitude $b_{01}(l,\tau-t_{d})$, Eq. (\ref{Eq7a}),
which does not take into account the spectrum narrowing of the photon. The
pulse area of $b_{01}(l,\tau-t_{d})$ (the area under the dashed line)
coincides with the area of the adiabatic part, $b_{0A}(l,\tau)$. Observing
only the tail of the photon envelope, one cannot distinguish the delayed
photon from the one that would reach the detector without absorber. However,
the leading edge of the photon envelope is split in two parts. The
non-adiabatic part is fast with a duration proportional to $1/\left(
T_{b}\Gamma)=1/(\alpha_{0}l\right)  $ and its amplitude equals unity at
$\tau=0$. This part originates from the spectrally broad component of the
photon. The thicker the sample, the shorter this part will be. The adiabatic
component is slow and propagates with group velocity $V_{g}$. Its leading edge
is broadened in time from $\tau=t_{d}-2\Delta_{\text{eff}}^{-1}$ to
$\tau=t_{d}+2\Delta_{\text{eff}}^{-1}$. The slow part of the photon does not
change appreciably in shape except the smoothening of the leading edge. This
is because for the chosen values of the parameters $\Gamma$, $\Omega$, and $T$
the effective width of the EIT hole is larger than the width of the photon
spectrum, i.e., $\Delta_{\text{eff}}>\Delta_{\text{ph}}$. For our numerical
example $\Delta_{\text{eff}}/\Delta_{\text{ph}}=6.9$. The delay time of the
photon, $t_{d}$, is 1.4 times longer than the lifetime of the excited state of
the source particle $\tau_{\text{life}}=\tau_{\text{ph}}/2$.

Since $\Delta_{\text{eff}}\approx\Omega^{2}/\Gamma\sqrt{T_{b}}$, one can
decrease the effective width of the hole and make it narrower than the
spectral width of the photon, $\Delta_{\text{eff}}<\Delta_{\text{ph}}$, by
reducing the coupling $\Omega$ or increasing the thickness $l$ of the sample
and hence increasing the effective thickness $T_{b}$. Then, it can be shown
that the shape of the slow photon amplitude reduces to almost a Gaussian.
However, its far tail for long times cannot decay more slowly than
$\exp(-\gamma_{m}t)$. Thus, if $\Delta_{\text{ph}}=\gamma_{m}$ and an EIT hole
is present, the far tail of the photon probability amplitude is always
indistinguishable from the case without absorber. This is because the induced
coherence of the states $g$ and $m$ has the same lifetime as the lifetime of
the photon probability amplitude, both being caused, for example, by the
spontaneous decay of the excited (for the source) and metastable (for the
absorber) states. The $g-m$ coherence (i.e., an antisymmetric superposition of
states $g$ and $m$) is at the origin of destructive interference. It can be
shown that this coherence reduces the absorption and as such supports the
photon propagation without absorption.

The indistinguishability of the tails of the slow photon and a photon that
does not propagate through an absorber can be removed if $\Delta_{\text{ph}%
}\gg\gamma$. In this case the slow part of the photon envelope transforms to
an almost Gaussian shape. Most part of its amplitude exceeds the probability
amplitude of the photon propagating without absorber. The area under its
envelope coincides with the area of $b_{01}(l,\tau-t_{d})$, i.e., with
$\theta_{A}(l)$ given by Eq. (\ref{Eq10}). Fig. 6(b) shows the time dependence
of the photon probability amplitude $b_{\text{tot}}(l,\tau)$ if $\Delta
_{\text{ph}}=\Gamma$ (bold line). The other parameters are the same as for
Fig. 6(a). Comparing the fast, nonadiabatic parts of the probability
amplitudes in Fig. 6(a) and 6(b), we find that they are almost the same in
spite of the difference of the halfwidth of the photon spectrum for (a) and
(b), being\ $\gamma_{m}$ and $\Gamma$, respectively. This is because the
spectrally broad component of the photon, responsible for the nonadiabatic
contribution, originates from the stepwise rise of the photon probability
amplitude at $t=0$ and does not depend on the spectral width of the radiation source.

Concluding this section we analyze the evolution of the symmetric and
antisymmetric components of the photon spectrum with distance in the EIT
medium. Their adiabatic $b_{s(A)}(l,\tau)$, $b_{a(A)}(l,\tau)$ and
nonadiabatic $b_{s(N)}(l,\tau)$, $b_{a(N)}(l,\tau)$ counterparts can be
calculated similarly to $b_{0A}(l,\tau)$ and $b_{0N}(l,\tau)$. Applying the
same procedure as before we obtain
\begin{equation}
b_{s,a(\text{tot)}}(l,\tau)=b_{s,a(A)}(l,\tau)+b_{s,a(N)}(l,\tau),
\label{Eq45}%
\end{equation}
where the adiabatic symmetric $b_{s(A)}(l,\tau)$ and antisymmetric
$b_{a(A)}(l,\tau)$ counterparts are
\begin{equation}
b_{s(A)}(l,\tau)=\left[  R_{+}(l,\tau)+R_{-}(l,\tau)\right]  /2, \label{Eq46}%
\end{equation}%
\begin{equation}
b_{a(A)}(l,\tau)=\left[  R_{+}(l,\tau)-R_{-}(l,\tau)\right]  /2. \label{Eq47}%
\end{equation}
The function $R_{\pm}(l,\tau)$ is defined as%
\begin{equation}
R_{\pm}(l,\tau)=\phi_{\pm}(l,\tau)\exp\left[  -T_{\text{eit}}\mp
\Delta_{\text{ph}}(\tau-t_{d})\right]  . \label{Eq48}%
\end{equation}
The nonadiabatic parts $b_{s(N)}(l,\tau)$, $b_{a(N)}(l,\tau)$ are calculated
in Sec. III. They are defined by the output field for the symmetric
$b_{s}(l,\tau)$ and antisymmetric $b_{a}(l,\tau)$ components given by Eqs.
(\ref{Eq26}),(\ref{Eq27}) for $\Delta_{\text{ph}}=\Gamma$ and by Eqs.
(\ref{Eq30}),(\ref{Eq31}) for $\Delta_{\text{ph}}=\gamma_{m}$. For the latter
case, $\gamma$ is substituted by $\gamma_{m}$ in Eqs. (\ref{Eq30}%
),(\ref{Eq31}). Fig. 7 shows the time dependence of the symmetric (a) and
antisymmetric (b) components of the photon probability amplitudes with a
narrow spectrum, $\Delta_{\text{ph}}=\gamma_{m}$. These plots clearly
demonstrate that the broad component of the photon spectrum is present only in
the antisymmetric time domain counterpart. Both counterparts are delayed due
to the reduced group velocity and they are smoothened. The discontinuities in
the time dependence of the time derivative for the symmetric part and of the
amplitude for the antisymmetric part are removed.

\section{Application of the EIT filtering}

The construction of a quantum network, consisting of quantum nodes and
interconnecting channels, is a challenge in quantum communication and
computation science. It is natural to use as nodes matter in the form of
individual atoms or atomic ensembles. Storage of a quantum state of light in
an EIT medium \cite{Fleischh} or in an extended ensemble of atoms with the
photon-echo technique \cite{Moiseev} looks very attractive. The experimental
realization of such protocols with a classical radiation field \cite{Phillips}
and faint laser pulses imitating a single photon radiation field \cite{Kroll}
showed the feasibility of storing and retrieval of a radiation field and
demonstrated a delayed self-interference for a single photon wave packet in
matter. If a single photon source of the first kind is used in these
protocols, the fidelity of the mapping of a quantum state of light to matter
can be quite low. To show this we introduce two quantum states%
\begin{equation}
\left\vert b_{s}\right\rangle =-i\sqrt{2}\sum_{\mathbf{k}}g_{\mathbf{k}}%
\frac{\Delta_{ph}\exp(-i\mathbf{k\cdot r}_{0})}{\nu_{k}^{2}+\Delta_{\text{ph}%
}^{2}}\left\vert 1_{\mathbf{k}}\right\rangle , \label{Eq51}%
\end{equation}
and%
\begin{equation}
\left\vert b_{a}\right\rangle =\sqrt{2}\sum_{\mathbf{k}}g_{\mathbf{k}}%
\frac{\nu_{k}\exp(-i\mathbf{k\cdot r}_{0})}{\nu_{k}^{2}+\Delta_{\text{ph}}%
^{2}}\left\vert 1_{\mathbf{k}}\right\rangle , \label{Eq52}%
\end{equation}
which correspond to the symmetric and antisymmetric parts of the photon
emitted in free space. The states are orthogonal and normalized: $\left\langle
b_{s}|b_{s}\right\rangle =\left\langle b_{a}|b_{a}\right\rangle =1$;
$\left\langle b_{a}|b_{s}\right\rangle =0$. This can be easily verified taking
into account that the Fock states $\left\vert 1_{\mathbf{k}}\right\rangle $
are orthogonal and the sum over $\mathbf{k}$ is replaced by an integral over
all frequency modes. The orthogonality of the states comes from the fact that
the symmetric part of a photon, $\left\vert b_{s}\right\rangle $, is an even
function of $\nu_{k}$ while the antisymmetric part, $\left\vert b_{a}%
\right\rangle $, is an odd function of $\nu_{k}$.

Formally we can represent the radiation state $\left\vert b\right\rangle $,
emitted by a single quantum particle in free space, [see Eq. (\ref{Eq1a}) in
Sec. II] as
\begin{equation}
\left\vert b\right\rangle =\frac{1}{\sqrt{2}}\left\vert b_{s}\right\rangle
+\frac{1}{\sqrt{2}}\left\vert b_{a}\right\rangle . \label{Eq50}%
\end{equation}
Such an expression allows us to make a strong statement about a
single photon interacting with a single atom or an ensemble of
atoms. First, it is evident that the photon emitted by a single
particle populates equally both states, $\left\vert
b_{s}\right\rangle $ and $\left\vert b_{a}\right\rangle $. In Sec.
III, we showed that the symmetric, $\left\vert b_{s}\right\rangle
$, and antisymmetric, $\left\vert b_{s}\right\rangle $, parts of
the radiation field interact differently with resonant atoms
(quantum absorbers in general). The antisymmetric part almost does
not interact with an atom while the symmetric part does. Also, it
is known (see, for example, Ref. \cite{Eberly}) that the
probability of excitation of an atom by a radiation field with a
temporal duration much shorter than the decay time of the atomic
coherence is defined by the pulse area. This area is zero for the
antisymmetric part of the photon and it has a finite value for the
symmetric part. Thus, a photon emitted by a photon gun of the
first kind has only a 50\% chance to interact with a target atom
and 50\% chance to miss the target atom. Such a poor score would
make the use of a single photon source of the first kind in
quantum computing or information storage unreliable. An
experimentalist performing a measurement cannot distinguish
whether missing a target is the result of quantum statistics
(i.e., the result of a small interaction probability) or a purely
quantum mechanical result originating from causality (i.e., the
result of pulling the trigger, which leads to populating the
noninteracting state $\left\vert b_{a}\right\rangle $). Therefore,
a qubit preparation in the form of matter excitation or any other
operation with such a qubit using a single photon source of the
first kind would become very uncertain and hence unreliable.

However, filtering a single photon through an EIT window may help
to make it much more reliable for quantum computing and
information storage. At the output of a thick EIT filter, the
broadband, $\left\vert b_{a}\right\rangle $, and narrowband,
$\left\vert b_{s}\right\rangle $, parts of the photon are well
separated in time. The broadband part can be removed by time
gating or a shutter and, for example, sent to an auxiliary
channel. Then, the probability amplitude of the photon at the
output of the EIT filter becomes bell-shaped (see Sec. IV).
Sending the removed broadband part of the photon to a detector
allows to purify the photon state. If the detector does not
"click", all probability amplitude is collected in a state that is
a wave packet with a
Gaussian envelope%
\begin{equation}
\left\vert b_{filtered}\right\rangle \approx\sum_{\mathbf{k}}\frac
{g_{\mathbf{k}}\Delta_{ph}}{\Delta_{ph}^{2}+\nu_{k}^{2}}e^{-i\mathbf{k\cdot
r}_{0}+i\nu_{k}t_{d}-\nu_{k}^{2}/\Delta_{eff}^{2}}\left\vert 1_{\mathbf{k}%
}\right\rangle . \label{Eq53}%
\end{equation}
If it clicks, no photon with a Gaussian shape is present. Such a detector in a
purification scheme can be omitted, since what comes out of the EIT filter
with appropriate time gating is always the pure state (\ref{Eq53}). However,
this auxiliary detector may help to conclude that the photon, emitted on
demand, failed to pass through the EIT filter and we have to repeat the operation.

One could argue that filtering the radiation emitted by a single photon source
reduces the total probability amplitude of the output photon while the
interaction probability of such a photon with a target atom (or atoms) remains
almost the same. So, what would be the gain and what would be the advantage of
the EIT filtering? The gain comes from the removal of the broadband part,
which does not interact with the target atom and hence produces a count at the
detector placed behind the target. As such it is a false count carrying no
information since the photon is assumed to be "stored" in the atom (atoms),
but it escapes the atom-field interaction due to the $\left\vert
b_{a}\right\rangle $-component. Therefore, the false counts can be considered
as noise in an information storage process. When applying filtering, this
noise would reduce to zero, while with no filtering we have to compare the
atom-field interaction probability with the quite high detection probability
($\sim$50\%) given by the broad spectral component of the photon $\left\vert
b_{a}\right\rangle $.

A further narrowing of the symmetric part of the photon in the case
$\Delta_{ph}>\Delta_{eit}$ works in the same way. We lose the "brightness" of
the source or the detection probability, but gain in spectral resolution.
Actually we do not lose the "useful" energy if our aim is to improve the
spectral resolution by sharpening the line associated to the source photon,
i.e., to make it more selective in the excitation of a particular target atom.
Selective excitation takes place if the spectral width of the source photon
equals the spectral width of the absorption line of the selected atom and if
their frequencies are in resonance. If our filter is designed such that these
requirements for the output radiation are fulfilled, only the "useful"
spectral content of the source photon is transmitted and the rest is
suppressed. As a result, the target atom interacts with such a photon with the
same probability as in the case of no filter and the "useful" brightness of
the source does not change. Otherwise, with no filter the inherent broad part
of the photon spectrum not interacting with the target atom (atoms) would
produce a click at the detector placed behind the target showing that the
atom-field interaction failed to happen.

To conclude this section, we consider the interaction of a Gaussian shape
photon with an ensemble of resonant atoms. We define the Gaussian wave packet
as
\begin{equation}
\left\vert b_{G}\right\rangle =\sum_{\mathbf{k}}\frac{2\sqrt{\pi}%
g_{\mathbf{k}}}{\Delta_{ph}}e^{-i\mathbf{k\cdot r}_{0}-\nu_{k}^{2}/\Delta
_{ph}^{2}}\left\vert 1_{\mathbf{k}}\right\rangle , \label{Eq54}%
\end{equation}
normalized such that it contains only one photon. The associated single photon
field $b_{G}(t)=\left\langle 0\right\vert E^{(+)}(\mathbf{r},t)\left\vert
b_{G}\right\rangle $ is $b_{G}(0,t)=b_{0G}(0,t)\exp(-i\omega_{0}t)$, where for
simplicity its amplitude $b_{0G}(0,t)=\exp(-\Delta_{ph}^{2}t^{2}/4)$ is
normalized to unity for $t=0$. The propagation of such a photon in a thick
absorber consisting of particles with the same resonant frequency $\omega_{0}$
is described by Eq. (\ref{Eq21}). We consider the case as in Sec. III, if the
absorption line in the absorber is inhomogeneously broadened with half width
$\Gamma$ [see Eq. (\ref{Eq29})]. To compare the absorption of a single photon
emitted by a source of the first kind with that for a photon with a Gaussian
envelope, we address the case of a narrow spectral width of a single photon:
$\Delta_{ph}\ll\Gamma$. In this case the integral Eq.(\ref{Eq21}) for a photon
with the Gaussian envelope, which describes its transmission through an
absorptive medium, can be calculated expanding the function $A_{\Gamma}(\nu)$,
Eq. (\ref{Eq29}), in a power series of $\nu$ near $\nu=0$. Keeping only three
terms of the expansion, we have%
\begin{equation}
A_{TL}(\nu)\approx\frac{\alpha_{0}}{\Gamma}\left(  1+i\frac{\nu}{\Gamma}%
-\frac{\nu^{2}}{\Gamma^{2}}\right)  . \label{Eq1B}%
\end{equation}
Then the integral is calculated analytically, i.e.,%
\begin{equation}
b_{0}(T,\tau)\approx\eta\exp\left[  -T-\eta^{2}\Delta_{ph}^{2}\left(
\tau+\frac{T}{\Gamma}\right)  ^{2}\right]  , \label{Eq3B}%
\end{equation}
where $T=\alpha_{0}l/\Gamma$, $\tau=t-l/c$ is the local time, $\eta
=1/\sqrt{1-fT}$, and $f=(\Delta_{ph}/\Gamma)^{2}$. This approximation is valid
if $fT<1$. The transmission function for a Gaussian photon, $U_{G}(T)$, can be
easily calculated%
\begin{equation}
U(T)=\frac{\sqrt{\pi}\eta}{\Delta_{ph}}e^{-2T}. \label{Eq4b}%
\end{equation}
For $\eta\sim1$, the deviation from the Beer's law is negligible, which shows
a strong atom-field interaction for such a photon. Thus, the state of a single
photon with a Gaussian envelope can be mapped into an atomic ensemble.
Recently, the mapping and retrieval of a classical field state in an ensemble
of resonant impurities in a solid was experimentally demonstrated
\cite{Manson}.

\section{Level mixing induced transparency for gamma radiation}

A similar analysis is applicable in the gamma domain, in particular to
reexamine the recent level crossing experiments in $^{57}$Fe
\cite{Coussement,Odeurs02}. In these references by means of M\"{o}ssbauer
spectroscopy \cite{MoesBook} the absorption of the 14.4 keV photons, emitted
by a radioactive single line source $^{57}$Co (CoRh), has been experimentally
studied in the natural mineral siderite (FeCO$_{3}$) containing $^{57}$Fe
nuclei. At the crossing of two hyperfine levels $m_{1}=-3/2$ and $m_{2}=+1/2$
in the first excited state of $^{57}$Fe \cite{Boolch94}, an appreciable
deficit of gamma-photon absorption was found. The experiments were performed
in frequency domain by Doppler shifting the frequency of the radiation source
and detecting the 14.4 keV photon, which passed through a thick resonant
absorber containing $^{57}$Fe. One can expect that time domain experiments
with TDCM could show a photon delay at the level crossing of $^{57}$Fe. We
assume that the fluctuating electron spin of Fe$^{2+}$ in FeCO$_{3}$ is the
dominant source of the line broadening for the transition $g-e$ \ in which the
ground state level $g$ is $m_{g}=-1/2$ and the excited state level $e$ is
$m_{1}=-3/2$. Our assumption is also supported by studies in Ref. \cite{Ok}.
Meanwhile, the transition $g-m$, where $m$ is the excited state level
$m_{2}=+1/2$ has mostly a natural broadening caused by spontaneous decay via
the emission of a gamma photon (into $4\pi$ angle) or via electron conversion
\cite{Ok}. If the excited levels $m_{1}=-3/2$ and $m_{2}=+1/2$ are coupled by
symmetry breaking interaction, then we have an EIT scheme with a coupling of
two excited levels having different line broadening mechanisms, one is due to
the electron spin fluctuations producing the linewidth $2\Gamma$ and the other
due to natural broadening, $2\gamma_{m}$, so that $\Gamma>\gamma_{m}$. This
EIT scheme will produce a slow photon shown in Fig. 6(a), where $\Delta
_{\text{ph}}=\gamma_{m}$. Another possibility for EIT is to apply an rf mixing
of two levels in the excited state of $^{57}$Fe \cite{ShakhOdeurs}.

The group velocity of the single photon wave-packet can be reduced many times
inside an absorber due to the steep dispersion in the EIT window. If the
propagation time of a slow photon in the absorber is much longer than the mean
dwell time between successive photons coming from the source, one can collect
several photons in the sample and then release them when required by means of
a sudden removal of all nuclei from resonance, producing a short burst of radiation.

\section{Discussion}

We have shown that a photon emitted by a single particle in free space
contains spectrally narrow and broad components with equal probability
amplitudes. The broad component has very low absorption in an ensemble of
two-level absorbers or a single atom. Therefore, a mapping of the state of
such a photon in the form of matter excitation has $\sim$50\% probability,
which is low for any realistic protocol of quantum information storage and
quantum computing. This is very different from what can be expected for a
single photon with a Gaussian time-envelope, which demonstrates a high
probability of atom-field interaction.

Filtering out the broad component of the photon with the help of an EIT medium
may produce a single photon that is (i) spectrally narrow and (ii) of Gaussian
shape (if $\Delta_{\text{ph}}>\Delta_{\text{eit}}$) in spite of an initially
highly asymmetric time dependence of the probability amplitude of the photon
coming from the source. Photon reshaping can be made at low cost in time and
energy. The narrow and broad spectral components of the photon are separated
in time and space because of different group velocities. The narrow part is
broadened in time (or spectrally narrowed) and delayed, while the broad part
transforms into a fast photon with a very short duration. The separation of
these components, for example by time gating, may produce a photon with a much
narrower spectrum, even with respect to the narrow part $\Delta_{\text{ph}}$
if $\Delta_{\text{ph}}>\Delta_{\text{eit}}$. The spectrum narrowing of the
narrow part takes place without appreciable amplitude probability loss. This
is because the time integrated probability amplitude of the photon, which is
analogous to the classical pulse area, is conserved and is not affected by the
spectrum narrowing of the photon. This area is just a parameter quantifying
the atom-field interaction. However, the time integrated probability of the
photon (its amplitude squared), which is analogous to the energy of a
classical pulse, decreases inversely proportional to the square root of the
effective thickness of the absorber. This decrease is much smaller than the
value given by Beer's law for monochromatic radiation.

Recently, another type of single photon sources based on spontaneous Raman
scattering of a laser field in an extended medium was reported \cite{Chou}%
-\cite{Kuzmich}. They are designed to implement the DLCZ protocol
\cite{DLCZ} for long-distance quantum communication with atomic
ensembles and linear optics . Spontaneous Raman scattering of a
low intensity laser field in an extended ensemble of three-level
atoms with states $\left\vert g\right\rangle $, $\left\vert
s\right\rangle $, $\left\vert e\right\rangle $, is capable to
produce a low intensity Stokes field in the forward direction with
an average photon number smaller than unity. In this protocol, the
laser field is strongly detuned from the resonant transition
$\left\vert g\right\rangle \rightarrow\left\vert e\right\rangle $
to insure low population of the excited state atoms, $\left\vert
e\right\rangle $. If the atoms are initially prepared in the
ground state $\left\vert g\right\rangle $, which the laser field
excites, the spin-wave (coherence of the ground states $\left\vert
g\right\rangle $ and $\left\vert s\right\rangle $ distributed
wave-like in the ensemble of three-level atoms) is excited along
with the scattering of one Raman photon into a wave packet
propagating in the forward direction. This event is probabilistic
and its success heralds the preparation of one quantum spin-wave
excitation stored in the atomic ensemble. It also demands the same
detuning of the Stokes photon from resonance as for the laser
field such that the Raman resonance condition for the two-quantum
process $\left\vert g\right\rangle \rightarrow\left\vert
e\right\rangle \rightarrow\left\vert s\right\rangle $ is
satisfied, i.e., the frequency difference of the laser field and
the scattered radiation equals the frequency of the transition
between the ground states $\left\vert g\right\rangle
\rightarrow\left\vert s\right\rangle $. If the Raman resonance
condition is not satisfied, with high probability no spin-wave but
rather the population of the ground state $\left\vert
s\right\rangle $ for a single atom (randomly localized spin
excitation) in the atomic ensemble is produced after spontaneous
Raman scattering on this atom with emission of a photon with a
resonant frequency $\omega_{es}=(E_{e}-E_{s})/\hbar$, where
$E_{e}$ and $E_{s}$ are the energies of states $\left\vert
e\right\rangle $ and $\left\vert s\right\rangle $. Therefore, this
kind of a single photon source satisfying the Raman resonance
condition needs a post-selection procedure. It implies the
generation of pairs of single photons, Stokes and anti-Stokes. One
photon of the pair (Stokes) is generated by the Raman scattering
described above. The laser producing this photon is named a "write
laser". It generates the photon and one quantum of the spin wave
excitation. Another photon of the pair (anti-Stokes) is generated
by a "read laser". Depending on the chosen scheme, it is applied
to the transition $\left\vert s\right\rangle \rightarrow\left\vert
e\right\rangle $ (in a three-level scheme) or $\left\vert
s\right\rangle \rightarrow \left\vert e^{\prime}\right\rangle $
(in a four-level scheme), where $\left\vert
e^{\prime}\right\rangle $ is an other excited state of the atom.
The read laser generates an anti-Stokes photon in a two-quantum
process $\left\vert s\right\rangle \rightarrow\left\vert
e\right\rangle \rightarrow \left\vert g\right\rangle $ or
$\left\vert s\right\rangle \rightarrow \left\vert
e^{\prime}\right\rangle \rightarrow\left\vert g\right\rangle $.
Detection of a single anti-Stokes photon after the read laser
pulse (post-selection) warrants the presence of a single quantum
of a spin-wave created by the write pulse and the generation of a
single photon state in the Stokes field satisfying the Raman
resonance condition. To reduce two-photon scattering in the Stokes
mode, the probability of a single photon emission, $n$, must be
small ($n<1$). Then the two-photon emission probability of the
Stokes field becomes even smaller $\sim n^{2}$. In this way a
correlated pair of single photon Stokes and anti-Stokes fields is
created. Stokes field generation in an atomic ensemble has
collectively enhanced the coupling to a certain optical mode due
to many-atom interference effects \cite{Duan}. This mode is
defined by Raman photon scattering into a wave packet propagating
forward along the write laser field and it is supported by the
spin-wave if the Raman resonance condition is satisfied. All Raman
scattering trajectories of the Stokes photon interfere
constructively in the forward direction. This is very similar to
the enhancement of single photon scattering in the forward
direction in an ensemble of two-level atoms considered in Sec. II
of our paper. The difference comes from the spectral content of
the produced photon. Due to the large detuning of the write laser
pulse and the post-selection of the Stokes photon satisfying the
Raman resonance condition, which warrants a single quantum
spin-wave excitation, the time envelope of such a photon is
completely defined by the spectral properties of the write pulse
and not by the lifetime of the excited state $\left\vert
e\right\rangle $. This was clearly shown for the Stokes photon
with photon number $n<1$ by Lukin's group in Ref. \cite{Eisman} .
For $n>1$, the spontaneous generation of the Stokes field changes
to stimulated emission, introducing a time-asymmetry of the
scattered field envelope. To reduce the spontaneous noise, it is
important to work with the write laser producing a Stokes photon
with $n\ll1$ (for example, $n=0.1-0.2$). In Refs.
\cite{Balic},\cite{Kuzmich} an additional phase matching condition
$\mathbf{k}_{w}+\mathbf{k}_{s}=\mathbf{k}_{r}+\mathbf{k}_{as}$,
where $\mathbf{k}_{w}$, $\mathbf{k}_{r}$ are wave vectors of the
write and read laser fields, and $\mathbf{k}_{s}$,
$\mathbf{k}_{as}$ are wave vectors of Stokes and anti-Stokes
photons, respectively, is applied to make the axis of the
write-read laser beams different from the axis of the scattered
Stokes-anti-Stokes photons. Any Stokes photon that does not
satisfy the condition of Raman resonance (or phase matching
condition) and that, on the contrary, has the resonant frequency
$\omega_{es}$ for the transition $e\rightarrow s$, will be
spectrally broadened since it is produced due to spontaneous decay
of the excited state $\left\vert e\right\rangle $ of a single
particle, and not collectively. In our paper we show that such a
photon or a photon produced by a single particle decaying in free
space vacuum modes interacts differently with two- and three-level
atoms. Therefore, the design of any quantum network should take
into account the spectral properties of such a single photon and
its interaction with atomic ensembles.

\section{Acknowledgements}

This work was supported by the FWO Vlaanderen, FNRS, and the IAP
program of the Belgian government. R.N.S. also acknowledges
support from Russian Fund for Basic Research (04-02-17082) and
CRDF CGP (RP1-2560-KA-03).

\newpage\begin{figure}[ptb]
\resizebox{0.75\textwidth}{!}{\includegraphics{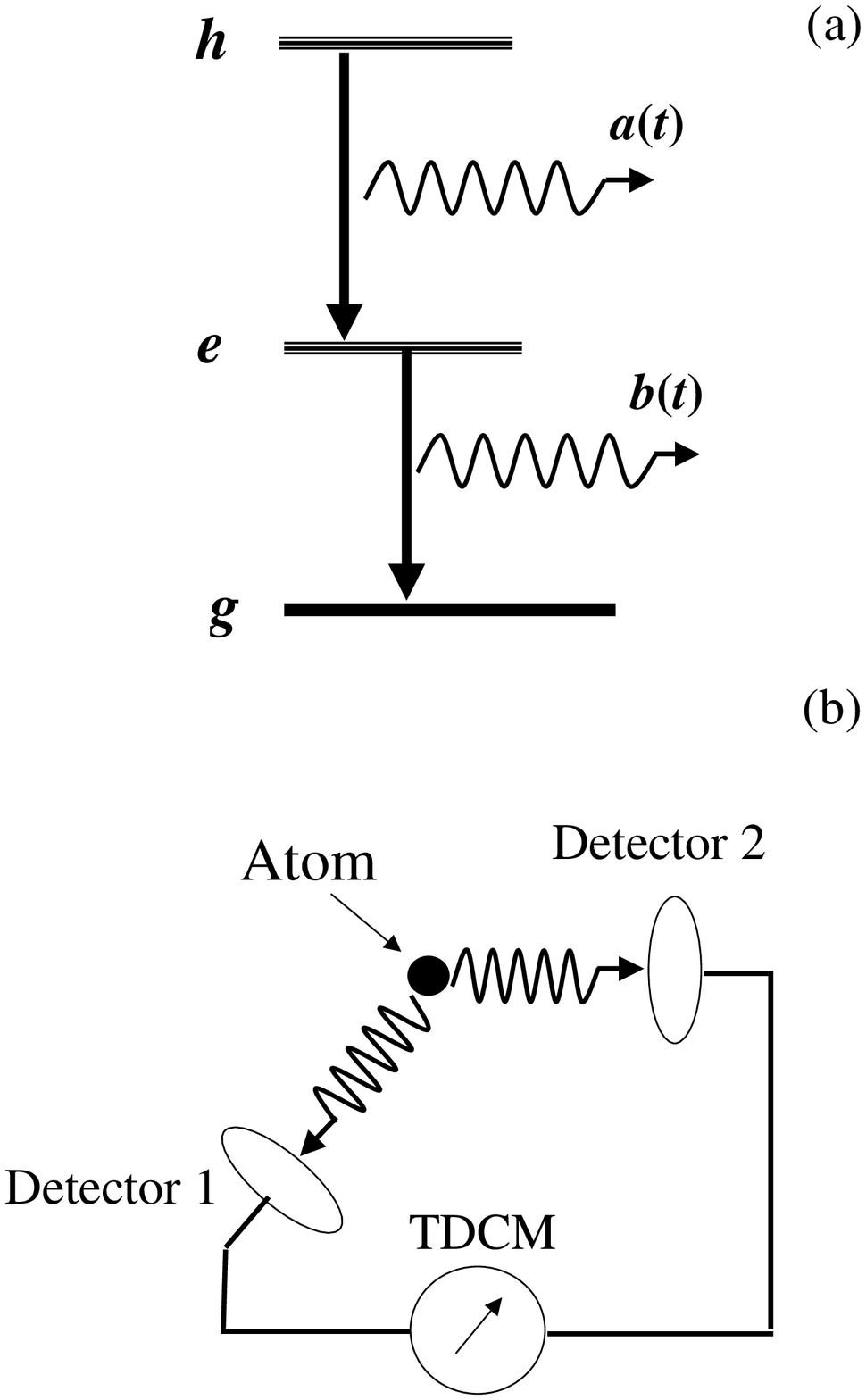}}\caption{(a)
Decay scheme of the excited particle emitting photons $a(t)$ and
$b(t)$ in the cascade $h\rightarrow e\rightarrow g$. (b) Detectors
1 and 2 detect photons $a(t)$ and $b(t)$, respectively. Detector 1
starts the clock when it detects a photon $a(t)$ and detector 2
stops the clock when the photon $b(t)$ is detected. The delay time
between the two counts is stored and after many such events the
detection probability of the photon $b(t)$ versus time is
reconstructed.}%
\label{fig:1}%
\end{figure}

\newpage\begin{figure}[ptb]
\resizebox{0.75\textwidth}{!}{\includegraphics{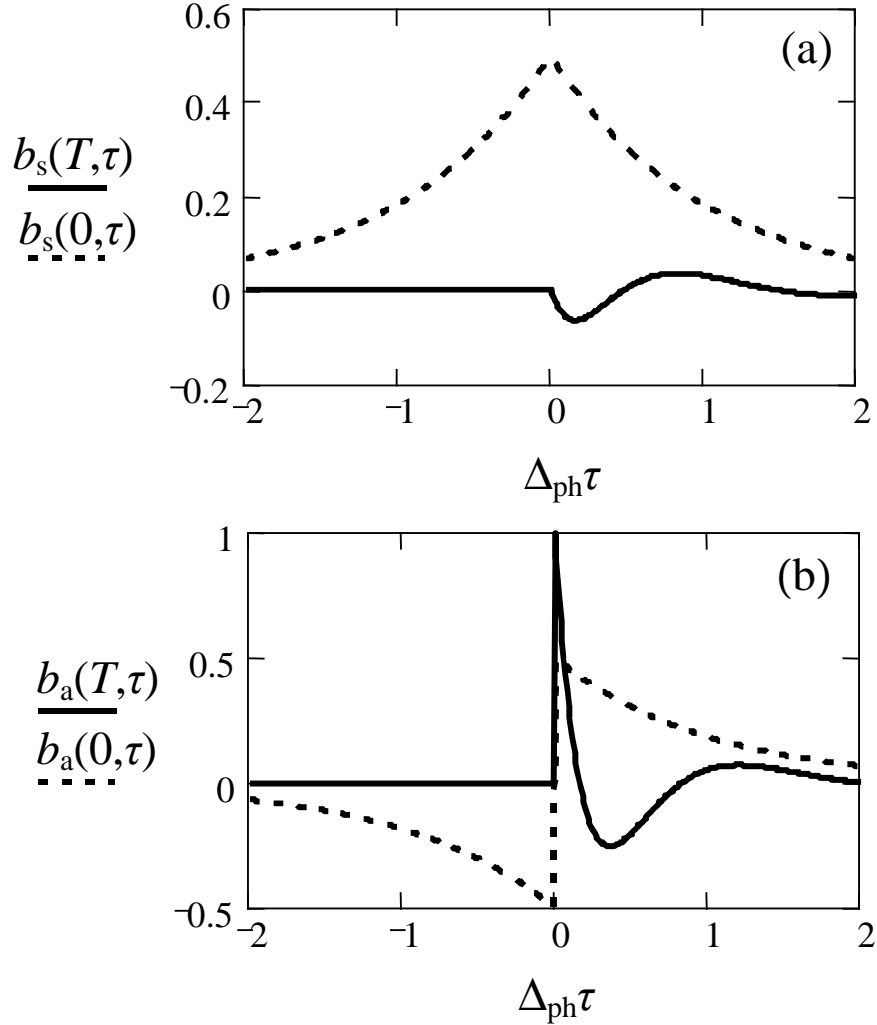}}\caption{Time
dependence of the probability amplitude of a photon, which is formally
presented as a sum of two parts. They are time domain counterparts of the
symmetric and antisymmetric components of the photon spectrum (see the text).
Plot (a) shows the time dependence of the symmetric component counterpart, and
(b) - of the antisymmetric component. Dashed line: no absorber, but a photon
is detected at distance $l$, so $t$ is substituted by $\tau=t-l/c$. Solid
line: with absorber of effective thickness $T=10$.}%
\label{fig:2}%
\end{figure}

\newpage\begin{figure}[ptb]
\resizebox{0.75\textwidth}{!}{\includegraphics{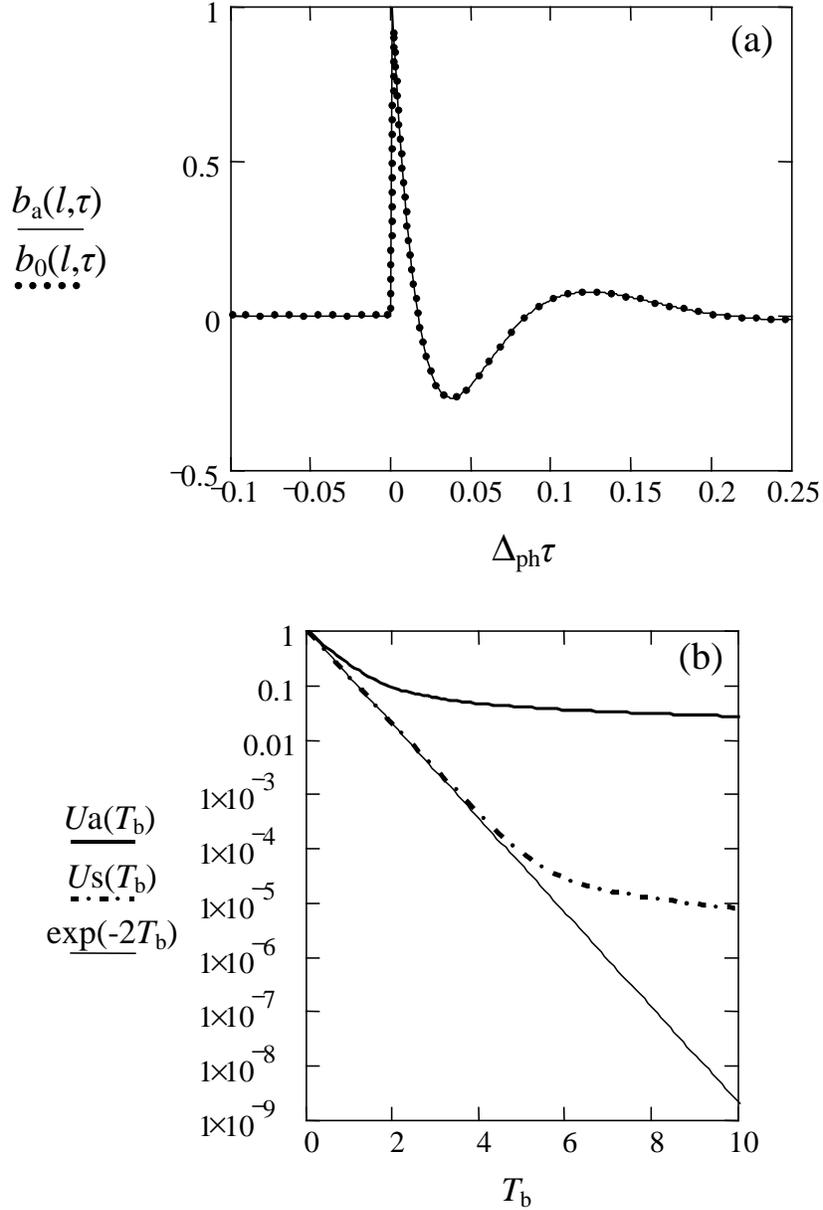}}\caption{(a) Time
dependence of the probability amplitude of the photon (dots) and its
antisymmetric part (solid line) at the output of a thick absorber with optical
thickness $\alpha_{0}l/\Gamma=10$. (b) Thickness dependence of the time
integrated intensity of the symmetric (dash-dot line) and antisymmetric (thick
solid line) parts of the photon. Both are normalized to half of the time
integrated intensity of the input photon $U_{0}(0)/2$. The thin solid line
shows the limit of Beer's law $exp(-2T_{b})$.}%
\label{fig:3}%
\end{figure}

\newpage\begin{figure}[ptb]
\resizebox{0.75\textwidth}{!}{\includegraphics{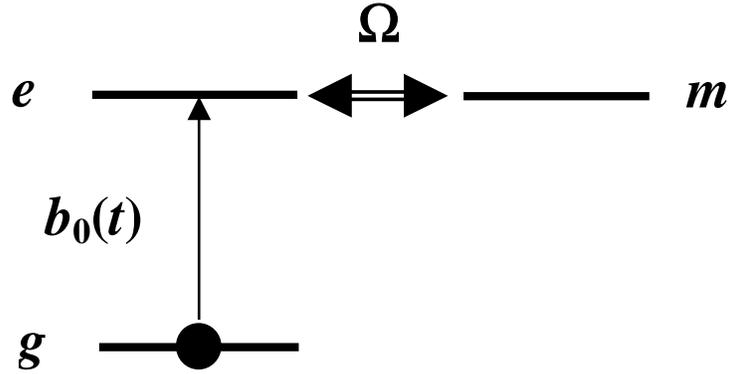}}\caption{Energy
diagram of the absorber whose excited state $e$ is coupled by $\Omega$ to a
metastable state $m$. Initially the particle is in the ground state $g$ and a
photon excites the transition $g \rightarrow e$.}%
\label{fig:4}%
\end{figure}

\newpage\begin{figure}[ptb]
\resizebox{0.75\textwidth}{!}{\includegraphics{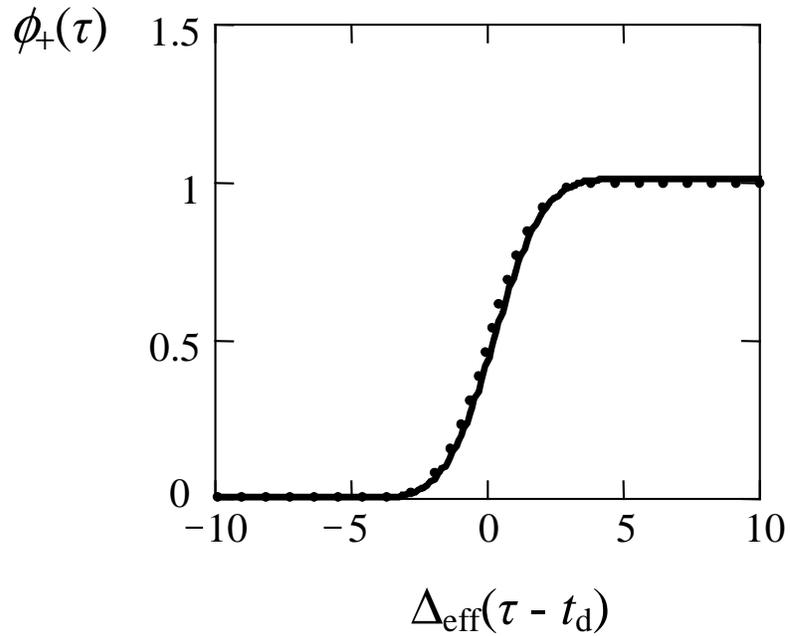}}\caption{Time
dependence of the function $\phi_{+}(l,\tau)$ for $\Delta_{ph}/\Delta
_{\text{eff}}$ equal to 0.1 (solid line) and 0 (dots).}%
\label{fig:5}%
\end{figure}

\newpage\begin{figure}[ptb]
\resizebox{0.53\textwidth}{!}{\includegraphics{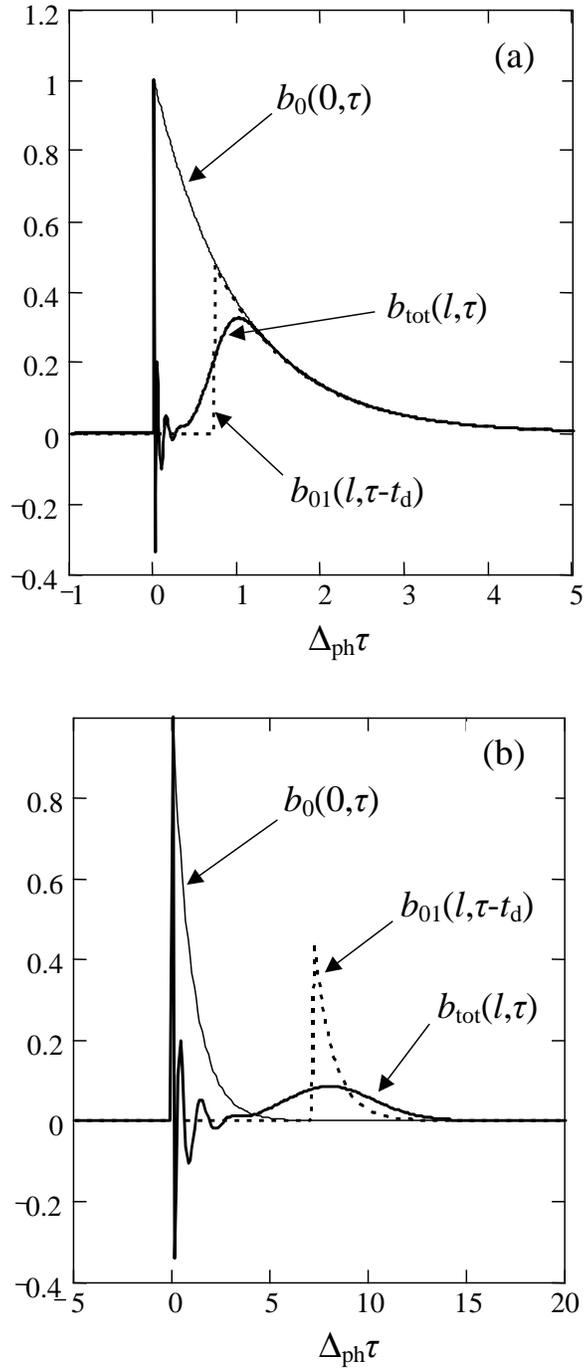}}\caption{Time
evolution of the probability amplitude of the photon, $b_{\text{tot}}(l,\tau
)$, transmitted through a sample of effective thickness $T_{b}=30$ (bold
line). The thin solid line shows the same evolution with no absorber,
$b_{0}(0,\tau)$. The dashed line shows the probability amplitude,
$b_{01}(l,\tau-t_{d})$, which does not take into account the spectrum
narrowing of the photon. The parameters of the absorber are $\Gamma
=10\gamma_{m}$ and $\Omega=2\Gamma$. The spectral halfwidth of the photon
$\Delta_{\text{ph}}$ is $\gamma_{m}$ (a) and $\Gamma$ (b).}%
\label{fig:6}%
\end{figure}

\newpage\begin{figure}[ptb]
\resizebox{0.75\textwidth}{!}{\includegraphics{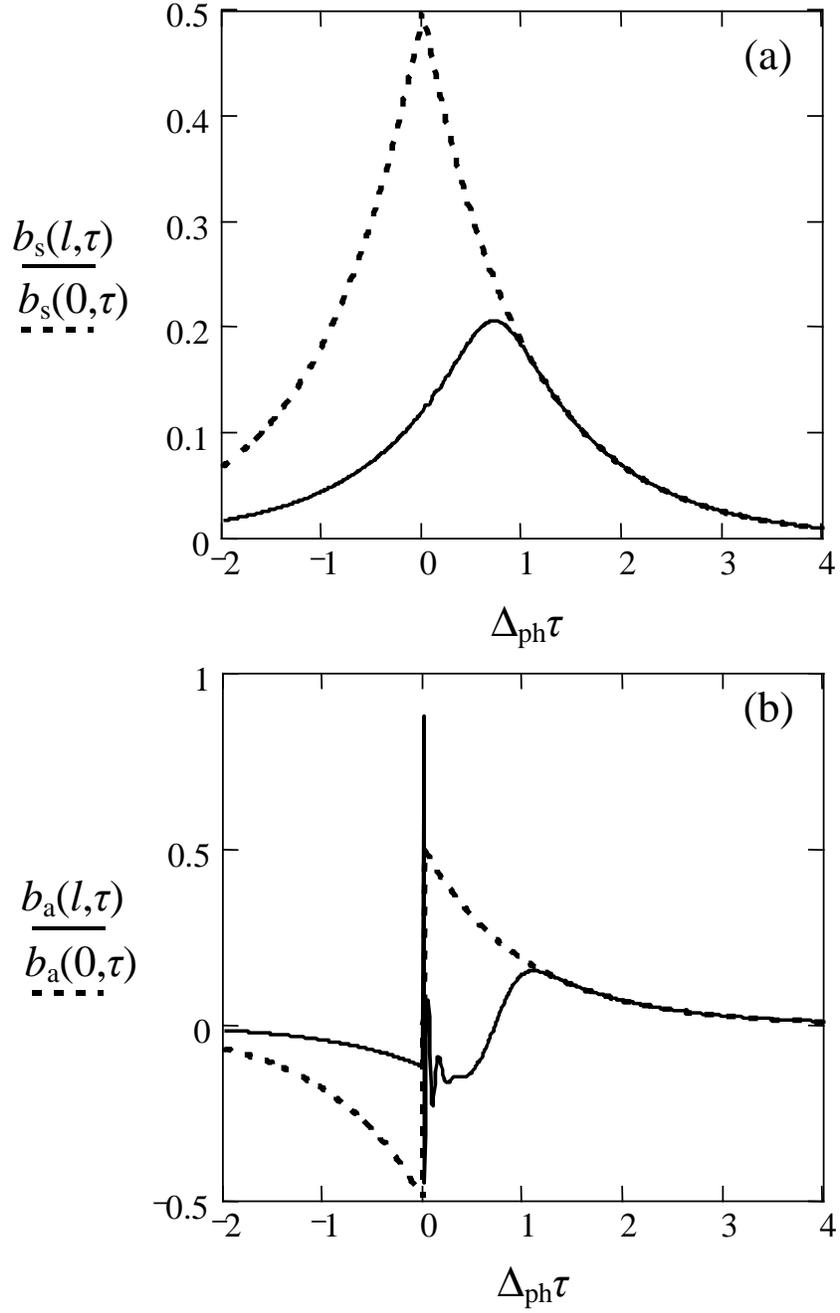}}\caption{Time
evolution of the probability amplitudes of the symmetric (a) and antisymmetric
(b) parts of the photon without absorber, $b_{s,a}(0,\tau)$ (dashed line), and
with absorber of effective thickness $T_{b}=30$, $b_{s,a}(l,\tau)$ (solid
line). The parameters of the absorber and photon are the same as in Fig.
6(a).}%
\label{fig:7}%
\end{figure}

\end{document}